\documentclass[runningheads]{llncs}
\usepackage{geometry}
\usepackage{cite}
\usepackage{censor}
\usepackage{textcomp}
\usepackage[utf8]{inputenc} % allow utf-8 input
\usepackage[T1]{fontenc}    % use 8-bit T1 fonts
\usepackage{hyperref}       % hyperlinks
\usepackage{url}            % simple URL typesetting
\usepackage{booktabs}       % professional-quality tables
\usepackage{amsfonts}       % blackboard math symbols
\usepackage{nicefrac}       % compact symbols for 1/2, etc.
\usepackage{microtype}      % microtypography
\usepackage{xcolor}         % colors
\usepackage{amsmath,amssymb,amsfonts}
\usepackage{algorithmic}
\usepackage{graphicx}
\usepackage{blindtext}
\usepackage{textcomp}
\usepackage[notransparent]{svg}
\usepackage{verbatim}
\usepackage{color,soul}
\usepackage{graphics}
\usepackage{listings}
\usepackage{multirow}
\usepackage{tabularx}
\usepackage{amssymb}
\usepackage{mwe}%

\def\BibTeX{{\rm B\kern-.05em{\sc i\kern-.025em b}\kern-.08em
    T\kern-.1667em\lower.7ex\hbox{E}\kern-.125emX}}

\usepackage{xcolor}
\usepackage{xspace}
\newif\ifdraft
\drafttrue

\ifdraft
  \newcommand{\mtnote}[1]{{\textcolor{orange}{ ***Matteo: #1 }}\xspace}
  \newcommand{\ozgurnote}[1]{{\textcolor{yellow}{ ***Matteo: #1 }}\xspace}
\else
 \newcommand{\mtnote}[1]{}
 \newcommand{\ozgurnote}[1]{}
\fi
    
\begin{document}

\title{Design and Implementation of an Analysis Pipeline for Heterogeneous Data}

\author{Arup Kumar Sarker \inst{1,3} \and
Aymen Alsaadi\inst{2}\and
Niranda	Perera\inst{5} \and
Mills Staylor\inst{1} \and
Gregor von Laszewski\inst{3} \and
Matteo	Turilli\inst{2,4} \and
Ozgur Ozan Kilic\inst{4} \and
Mikhail Titov\inst{4} \and
Andre Merzky\inst{2} \and
Shantenu Jha\inst{2,4} \and
Geoffrey Fox\inst{1,3}}
\authorrunning{Sarker, Alsaadi, Perera, Staylor, von Laszewski, et al.}
\institute{University of Virginia, Charlottesville, VA 22904 \email{\{djy8hg, qad5g, vxj6mb\}@virginia.edu} \and
Rutgers University, 57 US Highway 1. New Brunswick, NJ 08901-8554
\email{ \{aymen.alsaadi, matteo.turilli, shantenu.jha\}@rutgers.edu, andremerzky@gmail.com} \and
Biocomplexity Institute and Initiative, Town Center Four, 994 Research Park Boulevard Charlottesville, VA 22911
\email{laszewski@gmail.com} \and
Brookhaven National Laboratory, 98 Rochester St, Upton, NY 11973
\email{ \{okilic, mtitov\}@bnl.gov} \and 
Voltron Data, Mountain View, 650 Castro St, CA
\email{niranda.perera@gmail.com}}
%

\begin{comment}
\author{\IEEEauthorblockN{\censor{Given Name Surname}}
\IEEEauthorblockA{\censor{\textit{dept. name of organization (of Aff.)}} \\
\textit{\censor{name of organization (of Aff.)}}\\
\censor{City, Country} \\
\censor{email address or ORCID}}
\and
\IEEEauthorblockN{Given Name Surname}
\IEEEauthorblockA{\textit{dept. name of organization (of Aff.)} \\
\textit{name of organization (of Aff.)}\\
City, Country \\
email address or ORCID}
\and
\IEEEauthorblockN{Given Name Surname}
\IEEEauthorblockA{\textit{dept. name of organization (of Aff.)} \\
\textit{name of organization (of Aff.)}\\
City, Country \\
email address or ORCID}
\and
\IEEEauthorblockN{Given Name Surname}
\IEEEauthorblockA{\textit{dept. name of organization (of Aff.)} \\
\textit{name of organization (of Aff.)}\\
City, Country \\
email address or ORCID}
\and
\IEEEauthorblockN{\censor{Gregor von Laszewski}}
\IEEEauthorblockA{\textit{\censor{Biocomplexity Institute and Initiative}} \\
\textit{\censor{University of Virginia}}\\
\censor{Charlottesville, VA, 22911, USA} \\
\censor{laszewski@gmail.com} \\
\censor{https://orcid.org/0000-0001-9558-179X}}
\and
\IEEEauthorblockN{Given Name Surname}
\IEEEauthorblockA{\textit{dept. name of organization (of Aff.)} \\
\textit{name of organization (of Aff.)}\\
City, Country \\
email address or ORCID}
}
\end{comment}

\maketitle
\begin{abstract}
Managing and preparing complex data for deep learning, a prevalent approach in large-scale data science can be challenging. Data transfer for model training also presents difficulties, impacting scientific fields like genomics, climate modeling, and astronomy. A large-scale solution like Google Pathways with a distributed execution environment for deep learning models exists but is proprietary. Integrating existing open-source, scalable runtime tools and data frameworks on high-performance computing (HPC) platforms is crucial to address these challenges. 
%Combining RADICAL-Pilot, a Python runtime system, and Cylon, a distributed memory data parallel library,
%\mtnote{before saying that it is difficult to integrate RP and Cylon, should we say why we want to integrate them?} 
%opens a path for the open-source research community to use a heterogeneous distributed execution. But it is a complex process due to differences in architectures and technologies.
%Our goal is to create a seamless, integrated approach of data engineering, and deep learning frameworks with heterogeneous execution that is deployable on clouds, supercomputers, and HPC platforms, supporting heterogeneous systems with accelerators where Cylon and other data engineering and deep learning frameworks have a bridge to use heterogeneous execution. Thus, we integrated RADICAL-Pilot (RP) heterogeneous runtime system with Cylon parallel and distributed data framework (RP-Cylon) while executing Cylon as a task of RP. 
Our objective is to establish a smooth and unified method of combining data engineering and deep learning frameworks with diverse execution capabilities that can be deployed on various high-performance computing platforms, including cloud and supercomputers. We aim to support heterogeneous systems with accelerators, where Cylon and other data engineering and deep learning frameworks can utilize heterogeneous execution. To achieve this, we propose Radical-Cylon, a heterogeneous runtime system with a parallel and distributed data framework to execute Cylon as a task of Radical Pilot.
%The proposed design involves multiple masters and thousands of workers with task scheduling and resource allocation. Homogeneous tasks are executed in allocated nodes, combining multiple heterogeneous data pipelines and leveraging parallelism.
%We describe in detail the design and the development of RP-Cylon and the integration approach of Cylon tasks using RP to harness and utilize heterogeneous~\texttt{MPI-Communicators} across multiple nodes.
We thoroughly explain Radical-Cylon's design and development and the execution process of Cylon tasks using Radical Pilot. This approach enables the use of heterogeneous 
\texttt{MPI-Communicators} across multiple nodes.
Radical-Cylon achieves better performance than Bare-Metal Cylon with minimal and constant overhead. Radical-Cylon achieves (4$\sim$15)\% faster execution time than batch execution while performing similar join and sort operations with 35 million and 3.5 billion rows with the same resources.   
%The optimization of heterogeneous systems with technologies like MLIR is promising\mtnote{not sure what this sentence adds. If we keep it, MLIR should be expanded.}. 
The approach aims to excel in both scientific and engineering research HPC systems while demonstrating robust performance on cloud infrastructures. This dual capability fosters collaboration and innovation within the open-source scientific research community.
%\mtnote{We may want to speak about experiments and results in the abstract.}
\end{abstract}

\keywords{HPC \and BSP \and Cylon \and ETL \and MPI \and UCX \and RP \and BM \and SPMD \and MPMD}

\section{Introduction}\label{sec:introduction}

The exponential growth of data volume and complexity creates unprecedented challenges for the scientific community. With the increasing prevalence of sensors, internet-connected devices, and social media, the amount of data being generated is growing at an unprecedented rate. In addition, the complexity of scientific data is also increasing, with data coming from multiple sources often characterized by heterogeneity, high dimensionality, and complex relationships between variables, making it challenging to analyze data using traditional analysis tools. Even with the most advanced computing systems, processing massive datasets can be a significant bottleneck. Moreover, the sheer volume ofx data can make storing and transferring data across systems challenging. 

% and the total computing power in the world has quadrupled due to AI engine deployment

\begin{comment}
High-performance computing clusters have become more important for deep learning model analysis of big data sets. However, this is only one of several difficulties the data processing environment faces. Supporting data engineering for pre-processing, post-processing, communication, and system integration are additional difficulties. To boost user productivity and efficiency, data analytics solutions must be simple to integrate with already-existing frameworks in a variety of languages. This necessitates an effective and widely distributed data processing strategy, but many of the data analytics solutions available today are unable to satisfy all of these demands simultaneously. 
\end{comment}

These challenges have far-reaching implications across various scientific domains, including but not limited to genomics, climate modeling, physics simulations, and neuroscience. In genomics, for example, the amount of data generated from a single genome sequencing has grown exponentially, with individual genome sequencing now generating over 200GB of data~\cite{McKenna2010The}. Processing and analyzing such data using traditional methods can take months or even years. Similar to genomics, a single climate modeling simulation can generate vast amounts of data, with some simulations producing up to 10PB of data. Apart from huge growth, a data science survey conducted by Anaconda, indicates that a considerable amount of developer time (45\%) is dedicated to tasks like data exploration, preprocessing, and prototyping along with 33\% on deep learning tasks \cite{anaconda:online}. 

Analyzing and extracting insights from such massive data sets can be difficult, requiring a paradigm shift in how data analysis is performed, with the need for more efficient and scalable data analysis tools. Finding scalable solutions for processing data, such as large-scale simulations, modeling, and machine learning, is crucial for domain science, as it enables researchers to extract insights and knowledge from vast amounts of data efficiently. These solutions can directly impact reducing the time to solution and cost associated with data analysis by order of magnitudes, allowing scientists to focus on understanding complex systems that are not possible with smaller datasets and conduct more comprehensive and accurate analyses, leading to more robust and reliable results.

\begin{comment}
Modern data engineering frameworks must utilize the full potential of clusters to feed data to applications effectively. However, HPC frameworks like MPI, PGAS, or Slurm are not directly compatible with modern big data systems. Big data frameworks have made an effort to compete by offering comparable APIs, but doing so has a negative impact on performance because it requires switching between runtimes. Thus, there is a critical need for solutions that can leverage the processing power of HPC systems enabling analysis frameworks to accelerate their tasks while efficiently handling large datasets. 

Performing distributed operations quickly leads to faster results from data processing applications. Execution in parallel can improve the performance of aggregation procedures. Aggregations and grouping also need distributed execution to boost scalability given the volume of data created nowadays. Aggregation and grouping procedures are made more complex by parallel and dispersed execution, though. It is difficult but increases AI/ML user productivity to hide these intricacies from AI/ML users and give them access to a simple, recognizable abstraction. 
\end{comment}

Such solutions can be achieved by integrating existing scalable HPC runtime tools with data frameworks. Nevertheless, these solutions impose many research questions, such as performance optimization, programming models, efficient scalability, resource management, and utilization. Cylon\cite{abeykoon2020data} provides underlying frameworks for data engineering and deep learning applications to run on scalable HPC machines. However, it is not optimized for the efficient use of system resources and does not support a heterogeneous data pipeline. 
%To answer these questions, 
So, we first focus on a Python runtime engine that efficiently executes heterogeneous workloads of both executables/Python functions (non)MPI on a set of HPC machines. A task-based architecture, RADICAL-Pilot enable Cylon to interact and operate with different HPC platforms seamlessly, shielding Cylon from heterogeneous configurations of different HPC platforms. RADICAL-Pilot separates the resource management from the application layer, this would allow Cylon to run on any HPC resources without the need to refactor or rewrite the code, which reduces the development efforts, resulting in a loosely coupled integration. 

Furthermore, Cylon requires a heterogeneous runtime environment that constructs a private MPI/UCX/GLOO communicator for every Cylon task on HPC machines, which can be delivered using RADICAL-Pilot. Adding heterogeneous capabilities with RADICAL-Pilot and Cylon framework, which we call Radical-Cylon, can provide a powerful solution to these challenges as it enables the development of a unified system that can handle both compute and data-intensive workloads in an efficient and scalable manner. The experimental outcomes demonstrate that Radical-Cylon performs comparable to Bare-Metal (BM)-Cylon in strong and weak scaling scenarios involving join and sort operations. Radical-Cylon outperforms BM-Cylon in certain instances, particularly when the parallelism level reaches 512 or higher. For heterogeneous execution with strong and weak scaling of multiple tasks (e.g. combination of sort and join), Radical-Cylon showcases a performance improvement of (4$\sim$15)\% faster compared to batch execution with the same amount of resource utilization across all configurations with datasets of 35 million and 3.5 billion rows.

\section{Related Works} 
\label{sec:relatedWorks}

Recently, researchers in the field of distributed computing systems have been exploring various ways to improve the runtime of machine learning models. One such approach is the Ray framework\cite{moritz2018ray}, which proposes a new way of thinking about distributed computing for future AI applications. Although Ray is primarily intended for reinforcement learning scenarios, it has not yet been adopted for large-scale reinforcement learning systems due to some unresolved issues. However, Ray has had a significant impact and is considered to be positioned between K8S and deep learning frameworks, although it cannot replace them in these areas. Therefore, there is a need for an end-to-end framework that can optimize performance at every layer of the system, as the multiple components in each layer are tightly coupled, and the performance of each distributed model operation is affected if a single node or communication is not optimized.

The question of why we need a uniform distributed architecture arises, along with the bottlenecks of the current system. The answer is that we need an optimized system to reduce latency, but we also need to prove that the architecture is incremental and adaptive. To address this, Jeff Dean proposed a new concept of program execution patterns in his blog "Pathways: Next Generation AI Architectures." \cite{introduc61:online}. The framework is straightforward when considering Single Program Multiple Data (SPMD) criteria, but Multiple Program Multiple Data (MPMD) is composed of multiple SPMDs. Pathways \cite{barham2022pathways} departs from the traditional deep learning framework and design at a higher level to consider the best architecture for MPMD which aggregates all state-of-the-art Big Data frameworks. Hadoop\cite{introduc62:online} was the first generation of Big Data analytics and introduced the MapReduce programming model \cite{dean2008mapreduce}. However, Apache Spark\cite{zaharia2016apache} and Apache Flink\cite{carbone2015apache} surpassed Hadoop by providing faster and more user-friendly APIs. These advancements were made possible by hardware improvements that allowed for in-memory Big Data processing. Python Pandas Dataframes\cite{mckinney2011pandas} have emerged as the preferred data analytics tool among the data science and engineering community, despite being limited in performance and scalability. Dask Distributed and Modin are built on top of Pandas, providing distributed and generalized DataFrame abstractions, respectively. Later, CuDF emerged as a DataFrame abstraction that can be used for ETL pipelines on top of GPU hardware.   

OneFlow\cite{yuan2021oneflow} is a pioneering effort to revolutionize distributed data processing specifically within the realm of deep learning. 
%It introduces an actor-based framework that implements decentralized scheduling at runtime, offering a novel approach to managing computational resources. OneFlow boasts seamless support for parallel asynchronous dispatch of pathways, enabling efficient and flexible execution of tasks. Unlike making incremental changes to an existing framework, OneFlow was developed from scratch, allowing for a fresh design perspective that minimizes dependencies on pre-existing infrastructure. 
The authors of OneFlow assert that their framework has the potential to replace Plaque, a component within the Pathways system. 
%However, as Plaque has not been released for public use, the specific mechanisms by which OneFlow would serve as a replacement are not yet fully elucidated. 
%Nonetheless, OneFlow's innovative architecture and decentralized scheduling hold promise for improving distributed data processing in deep learning applications. 
ZeroMQ \cite{zmq1:online} is a high-performance asynchronous messaging library that provides communication between different applications or parts of an application over transport protocols. It can be a potential alternative to MPI. Arkouda \cite{arkouda:online} supports ZeroMQ-based communication protocol but has limitations in supporting heterogeneous data pipelines. 
Parsl\cite{babuji2019parsl} is a parallel scripting library designed to enhance Python with straightforward, scalable, and adaptable elements for representing parallelism. RP employs specifically to create a dynamic dependency graph of components.

\section{Design and Implementation}
\label{sec:design}

A unique integration approach of Cylon with the RADICAL-Pilot runtime system via their native Application Programming Interface (API) is proposed on Radical-Cylon. In the core system, RADICAL-Pilot is used as the distributed runtime for managing the execution of Cylon tasks. Importantly, we consider both systems ‘as they are’ in a loosely coupled design without the need to implement an integration plugin while exposing both system capabilities via RADICAL-Pilot API. This approach allows Cylon to benefit from RADICAL-Pilot heterogeneous runtime capabilities, specifically the capabilities to construct and deliver MPI communicators without modifying Cylon tasks. Additionally, the proposed design offers a flexible and adaptable framework for developing and deploying data-intensive applications on various HPC platforms.

\subsection{Radical-Pilot (RP)}

RADICAL-Pilot is a flexible and scalable runtime system designed to support the execution of large-scale applications on HPC leadership-class platforms. RADICAL-Pilot enables the execution of concurrent and heterogeneous workloads on various HPC resources. Further, RADICAL-Pilot offers the capabilities to manage the execution of (non)MPI single/multi-thread/core/node executables and functions efficiently. 

\begin{figure}[htpb]
    \begin{center}
        \includesvg[width=0.7\linewidth]{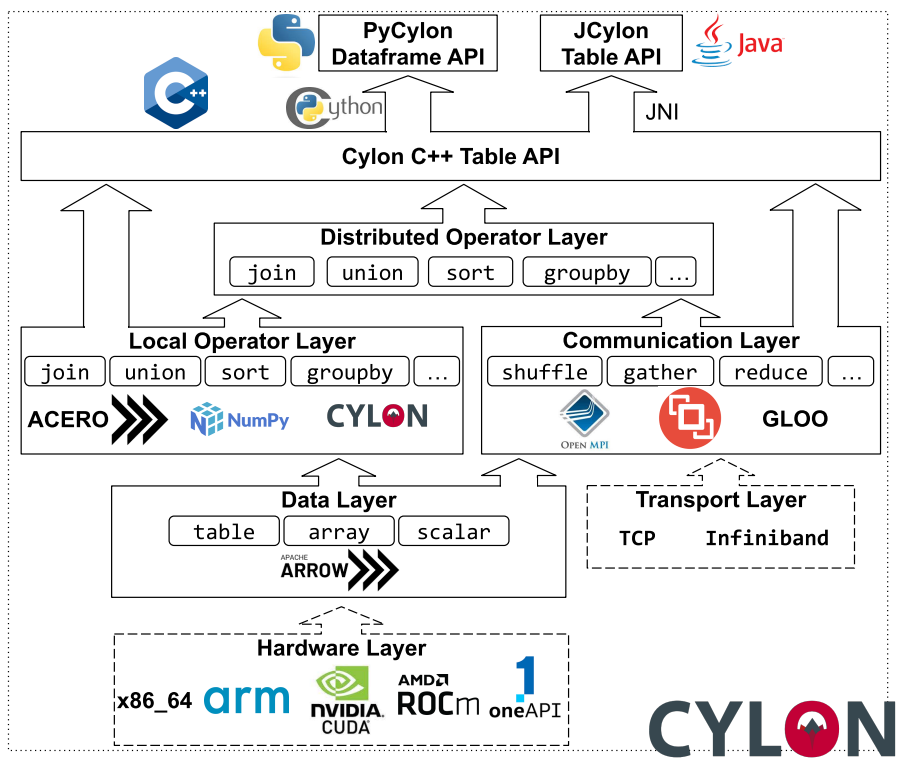}
        \end{center}
        \caption{Cylon Layered Architecture. From the bottom-up view, the Hardware layer is compatible with vendor-based or open-sourced transport layer \cite{perera2023depth}}
    \label{fig:cylon-arch}
\end{figure}

RADICAL-Pilot consists of three main components: The \emph{PilotManager},  \emph{TaskManager}, and \emph{RemoteAgent}. Both \emph{PilotManager} and \emph{TaskManager} run on user resources such as local computer or login/compute node of a cluster or HPC platforms, while the \emph{RemoteAgent} resides on the compute resources~\cite{merzky2022radical}.

The \emph{PilotManager} is responsible for managing the lifecycle of the pilot, which is a placeholder for resources on HPC systems. The \emph{TaskManager} is responsible for managing the lifecycle of tasks, which represent an application such as a function or executable that runs on the pilot's available resources. The \emph{RemoteAgent} is responsible for preparing the execution environment and starting the pilot to execute the tasks on the remote resources.

RADICAL-Pilot enables efficient scheduling, placement, and launching of independent tasks across multiple compute nodes. Leveraging the pilot abstraction model, RADICAL-Pilot has demonstrated the ability to concurrently execute up to one million tasks across one thousand nodes with minimal overheads~\cite{merzky2022radical}.

\subsection{Cylon}
Cylon represents a profound evolution in the realm of data engineering, offering a comprehensive toolkit that seamlessly connects AI/ML(with PyTorch~\cite{pytorch2019} and TensorFlow~\cite{tensorflow2015}) systems with data processing pipelines \cite{widanage2020high}.
Cylon's overarching vision is rooted in the fusion of data engineering and AI/ML, as exemplified by its ability to effortlessly interact with a spectrum of data structures and systems and optimize ETL performance. 
%It seamlessly integrates with prominent frameworks like PyTorch~\cite{pytorch2019} and TensorFlow~\cite{tensorflow2015}, amplifying their capabilities and propelling data-driven insights to new heights. 
%Furthermore, Cylon extends its reach beyond the confines of a library, evolving into a dynamic framework that underpins both ETL processes and the distributed modeling of AI/ML workloads. 

At the heart of Cylon's architecture, there is a core framework, wielding a sophisticated table abstraction to represent structured data in Fig.~\ref{fig:cylon-arch}. This abstraction empowers individual ranks or processes to collectively handle partitions of data, fostering a sense of unity and collaboration despite distributed computing challenges. Cylon's arsenal of "local operators" execute operations solely on locally accessible data, while "distributed operators" harness network capabilities to execute complex tasks that necessitate inter-process communication.

\begin{figure}[htpb]
    \begin{center}
        \includesvg[width= 0.7\linewidth]{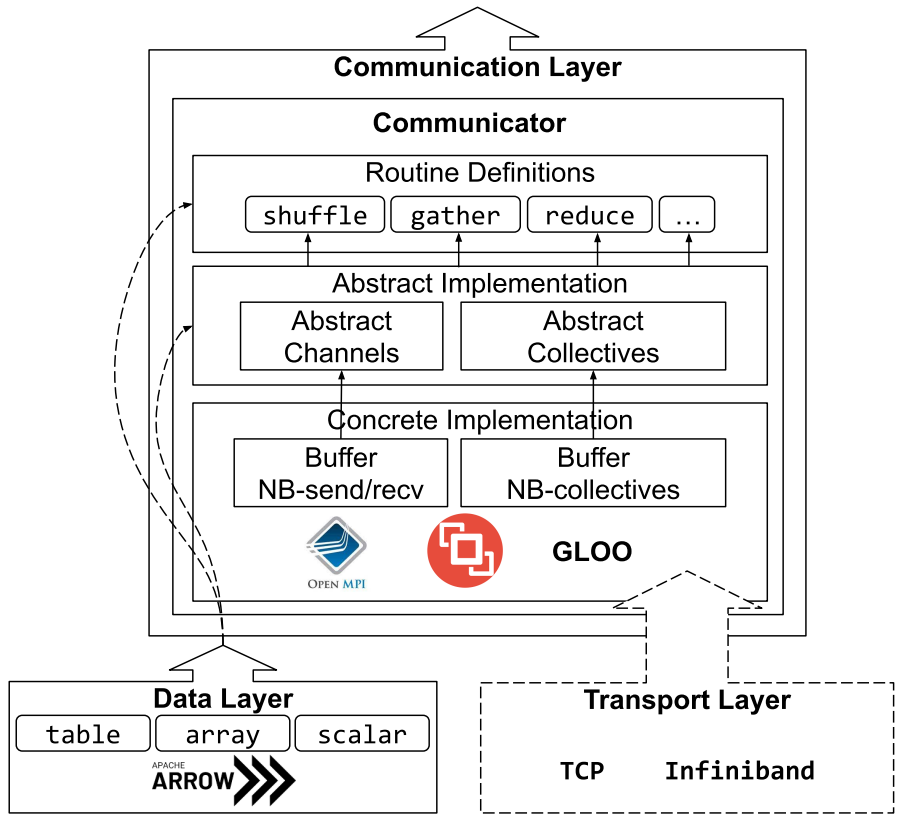}
        \end{center}
        \caption{Cylon Communicator Model. It has cross-platform support of Open-MPI, GLOO and UCX \cite{perera2023depth}}
    \label{fig:comm-model}
\end{figure}

To mitigate the complexities of distributed programming, Cylon orchestrates network-level operations that transpire atop communication protocols (Fig.~\ref{fig:comm-model}) like TCP or Infiniband. This allows multiple communication abstraction frameworks, e.g., MPI, UCX \cite{shamis2015ucx}, and GLOO\cite{gloo:online} for heterogenous data transmission \cite{shan2022hybrid, perera2023supercharging}. This strategic approach of channel abstraction elevates the efficiency of Cylon's operations(e.g. shuffle, gather, reduce, etc.), enabling seamless communication between processes while harmonizing performance across diverse hardware environments.

With Apache Arrow's Columnar Format as its foundation, Cylon's data model aligns seamlessly with a myriad of open-source frameworks and libraries. This interoperability ensures a harmonious coexistence within the larger data ecosystem, facilitating the exchange of data and insights between different platforms. In this paper, we embark on an exploration of Cylon's rich tapestry, dissecting its core components and intricate layers as an abstraction of RADICAL-Pilot. From data models and operators to communication and transport layers, Cylon's architecture emerges as a core component of distributed high-performance frameworks to reshape the future of data engineering and AI/ML integration.

\subsection{Design}

Cylon and RADICAL-Pilot are two isolated systems offering different functionalities and capabilities. The integration design advocates the loosely coupled approach where both systems work independently of each other, with minimal dependencies and interactions, while benefiting from each other's capabilities.

The integrated design of RADICAL-Pilot and Cylon is shown in Fig.~\ref{fig:cylon_rp_design} where Cylon is plugged as a top-level component to send different types of Cylon tasks (functions or executables) to RADICAL-Pilot to execute on HPC resources. The main communication point between both systems is their native APIs, as both systems offer flexible and simple Python-based interfaces. 

Cylon and RADICAL-Pilot loosely coupled integration can be easily scaled out, expanded, or contracted to meet changing demands~\cite{de_dynamio_2007}. Flexibility-wise, both systems are developing rapidly and might introduce new fundamental changes in the design or implementation, such as adding or removing new system components. Further, any changes in both systems do not necessarily require changes to the other system's components and would not affect the existing integration as there are no direct dependencies between the integrated systems. From a fault tolerance perspective, the integration approach of Cylon and RADICAL-Pilot is more resilient, as failures in one system or component do not affect the entire system. Failure of any component can be isolated and contained, allowing the rest of the system to continue receiving and executing tasks.

\subsection{Implementation}

\begin{figure}[htbp]
\begin{center}
\includesvg[width=0.55\linewidth]{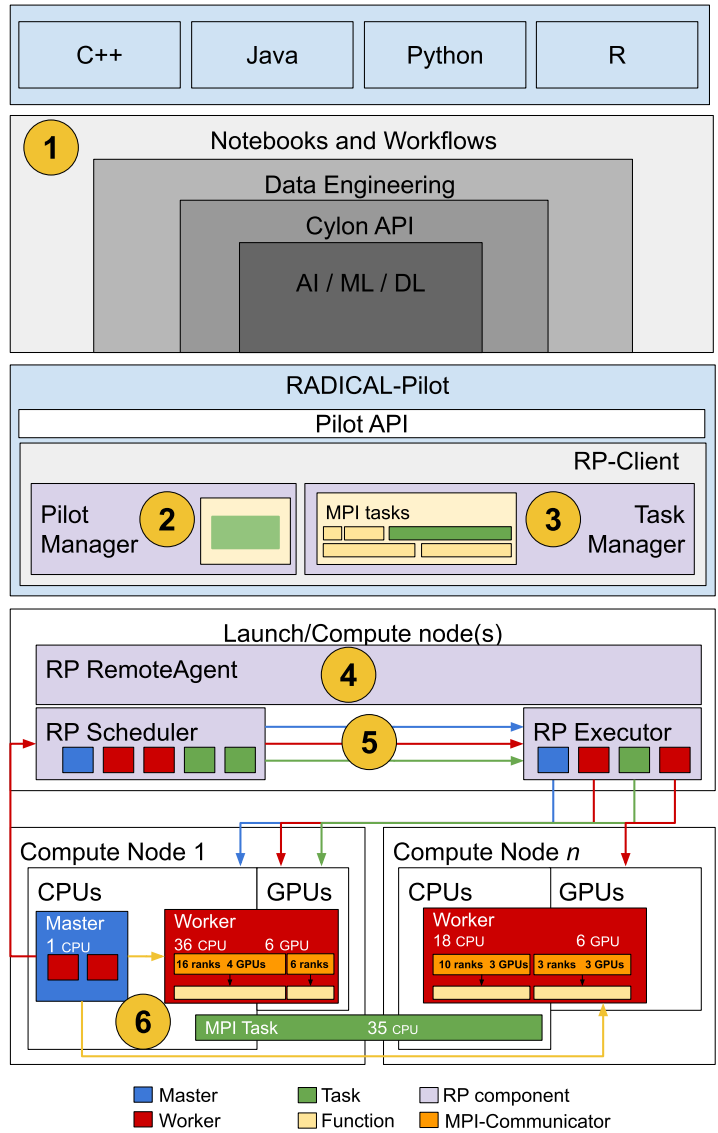}
\end{center}
\caption{Radical-Cylon Architecture. A modular design with dependent components. Segregated independent module with top-down flow from cross-platform to hardware resources. }
\label{fig:cylon_rp_design}
\end{figure}

We implemented Radical-Cylon as a single system enabling communication between the two systems via their Python APIs, as shown in Fig.~\ref{fig:cylon_rp_design}. The implementations expose RADICAL-Pilot API as a main interface to specify, interact, and execute Cylon tasks on multiple HPC platforms. Further, each Cylon task is represented as a ~\texttt{RadicalPilot.TaskDescription} class with their resource requirements, such as the number of CPUs, GPUs, and memory.

Once the Radical-Cylon starts~(Fig.\ref{fig:cylon_rp_design}-1), RADICAL-Pilot instructs the~\texttt{Pilot\-Manager} to create the~\texttt{Pilot} object with the required number of resources~(Fig.\ref{fig:cylon_rp_design}-2). Further, RADICAL-Pilot creates the~\texttt{TaskManager} and submits Cylon tasks to the~\texttt{TaskManager} to be executed on the remote resources~(Fig.\ref{fig:cylon_rp_design}-3). Synchronously, once the pilot resources are acquired from the HPC resource manager, RADICAL-Pilot starts the~\texttt{RemoteAgent} on the acquired resources~(Fig.\ref{fig:cylon_rp_design}-4). Once the~\texttt{RemoteAgent} is bootstrapped and ready, it starts the RAPTOR subsystem~(Fig.\ref{fig:cylon_rp_design}-5), which is an abstraction of the Master-Worker MPI paradigm. RAPTOR implementation is based on~\texttt{mpi4py}~\cite{dalcin2005mpi4py} and can concurrently execute heterogeneous MPI/non-MPI functions across multiple nodes. Unlike other pilot systems, RADICAL-Pilot and via RAPTOR offer the capabilities of constructing private MPI communicators of different sizes during the runtime, which Cylon tasks require.

Once all RAPTOR master(s) and worker(s) start, the master(s) receives the Cylon tasks from the~\texttt{RemoteAgent} scheduler and distributes them across the workers to be executed. When the worker receives Cylon tasks, it isolates a set of
 ~\texttt{MPI-Ranks} based on the resource requirements of the Cylon task and groups them to construct a private~\texttt{MPI-Communincator} and deliver it to the task during runtime~(Fig.\ref{fig:cylon_rp_design}-6). Finally, once all of Cylon's tasks finish execution, the master collects the results of the tasks and sends them back to the~\texttt{TaskManager}.

\begin{figure}[tbp]
\begin{center}
\includesvg[width= 0.8\linewidth]{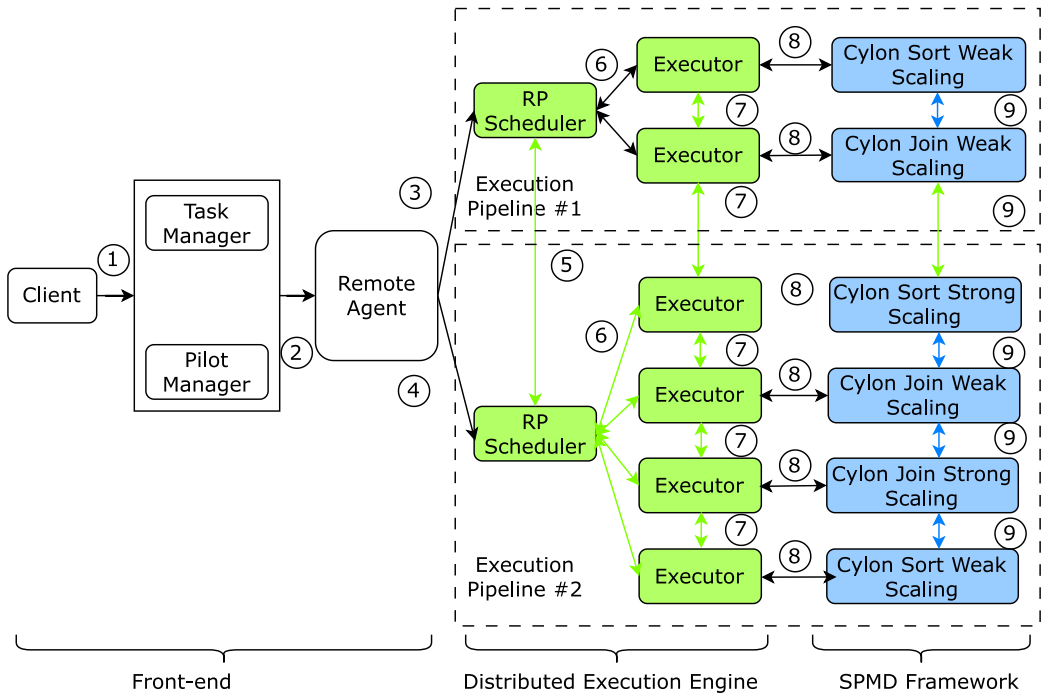}
\end{center}
\caption{Heterogeneous Execution with Control and Data Flow. The execution pipeline uses a separate SPMD framework for underlying tasks.}
\label{fig:radical_cylon_design}
\end{figure}

A bird's-eye view of the Radical-Cylon system is shown in Fig.~\ref{fig:radical_cylon_design}, with in-depth components and data flow. In the initial step (Step 1), when a user intends to execute a traced program (MPMD) comprising multiple computations (SPMD), they employ the Radical-Cylon system by invoking the RP-Client. Moving to Step 2, the Pilot Manager assigns virtual devices for computations not previously executed and registers these computations with the Resource Manager.

Subsequently, in Step 3, the client activates the background server to execute instructions for the pilot manager, incorporating considerations for network connections between devices and data routing operations among various computations. If the virtual device for a program remains unchanged, the generated representation can be swiftly reused; however, if the Resource Manager alters a program's virtual device, recompilation is necessary. These three steps collectively form the front end of Radical-Cylon. Remote Agent creates multiple execution pipelines with two persistent daemons – a scheduler and an executor – capable of communicating (Steps 5, 6, 7) to achieve distributed coordination, constituting the control plane communication. 

The executor invokes Cylon data engineering frameworks (Step 8) to perform local sorting or joining, or data plane communication as indicated in Step 9 (primarily cluster communication involving shuffle or gather operations). Notably, the communication between the data plane employs the same communication framework, with the former depicted by a blue arrow indicating higher bandwidth, and the latter represented by a green arrow indicating lower bandwidth.

\section{Experiments}

Table ~\ref{tab:exp_table} shows the setup of our experiments. We use UVA Rivanna HPC~\cite{rivanna_2023} and ORNL-Summit\cite{summit_2023} to set up weak and strong scalability experiments. On Rivanna, we use the \texttt{parallel} queue with 37 cores per node and a maximum of 14 nodes, and on ORNL-Summit, we use a maximum of 64 nodes with 42 cores per node.  We evaluate the efficiency of Radical-Cylon and compare it to Bare-Metal Cylon (BM-Cylon) while executing Cylon join and sort operations with single pipeline execution. We measure two metrics: Total Execution Time and Radical-Cylon overheads. The Total Execution Time represents the total time Radical-Cylon spent executing the join and sort tasks on the computing resources with N ranks. The Overheads represent the time taken by Radical-Cylon (mainly RP) to (i) deserialize the task object and (ii) construct the \texttt{MPI-Communicator} with N ranks and deliver it to the tasks. Each join and sort task takes N ranks with a maximum of 35 million rows per rank for weak scaling, and 3.5 billion rows are divided into N ranks for strong scaling.
Collectively, experiments~\ref{subsec:exp-join}, ~\ref{subsec:exp-sort}  and ~\ref{subsec:exp-homogeneous-cylon} allow us to study and evaluate the scalability performance of Radical-Cylon while comparing it to BM-Cylon on Rivanna and batch execution of BM-Cylon on ORNL-Summit with the setup of multiple configurations.

%For each task, we ensured the allocation of optimal resources per node. %In the parallel partition of Rivanna, we harnessed up to 37 cores per %node and utilized a maximum of 35 million rows per worker for the %sorting and joining weak scaling.

\begin{table*}
	\caption{Experiments Setup on UVA.Rivanna and ORNL.summit. WS/SS = weak/strong 
         scaling; M=Million; B=Billion; rank=1 physical core; RN=Rivanna Nodes; SN=Summit Nodes; RC=Rivanna CPU; SC=Summit CPU}
	\label{tab:exp_table}
	\centering
	\begin{tabular}{lllclcc}
		\toprule
		ID                                  &
		Experiment Type                     &
		RN                                  &
            SN                                  &
		Rows Size                           &
		RCs(ranks)                          &
            SCs(ranks)                          \\
        \midrule
		A                                   &
        Join Operation WS/SS                &
        $4-14$                              &
        $2-64$                              &
        [35M | 3.5B]                        &
        \#nodes $\times$ 37                 &
        \#nodes $\times$ 42                 \\
        \midrule
		B                                   &
        Sort Operation WS/SS                &
        $4-14$                              &
        $2-64$                              &
        [35M | 3.5B]                        &
        \#nodes $\times$ 37                 &
        \#nodes $\times$ 42                 \\
        \midrule
		C                                   &
        Heterogeneous WS/SS                   &
        $ $                              &
        $2-64$                              &
        [35M | 3.5B]                         &
                                             &
        \#nodes $\times$ 42                 \\
		\bottomrule
	\end{tabular}
\end{table*}

\begin{table}
	\centering
	\caption{Radical-Cylon((RP-Cylon)) Execution Time and Overheads of Strong and Weak Scaling from Experiment A (Join Operations) and B (Sort Operation) on Rivanna. %\mtnote{We have to add the data of Summit, otherwise the reviewers will complain that the presentation is partial. As we are very tight with space, you could span the table on two columns (use the star) and then put weak and strong as columns instead of as rows. After that, add a new table for Summit as the parallelism is different.}
 }
	\label{tab:a-b-exp_table}
	\begin{tabular}{llcr @{\hspace{1\tabcolsep}} lr @{\hspace{1\tabcolsep}} l}
		\toprule
		                   &
		               &
		                     &
		\multicolumn{2}{c}{Execution Time}         &
	    \multicolumn{2}{c}{Overheads}              \\
		Operation             &
		Scaling               &
		Parallelism                     &
		\multicolumn{2}{c}{time (seconds)}                    &
	    \multicolumn{2}{c}{(tasks/second)}                    \\
		\midrule
		\multirow{8}{*}{Join} &
		\multirow{4}{*}{Weak} &
		148                       &
		215.64 & $\pm4.35$     &
		2.9 & $\pm0.1$          \\
		&
		                     &
		222                     &
		226.12 & $\pm2.59$   &
		2.3 & $\pm0.4$        \\
		&
		                   &
		296                     &
		237.01 & $\pm2.96$    &
		2.8 & $\pm0.8$       \\
		                    & &
		370                     &
		239.87 & $\pm3.41$      &
		2.5 & $\pm0.8$        \\
		                    & &
		444                     &
		241.59 & $\pm2.76$      &
		2.9 & $\pm0.4$        \\
		                    & &
		518                     &
		253.66 & $\pm1.53$      &
		3.2 & $\pm0.6$        \\

		\cmidrule{3-7}
		&
		\multirow{4}{*}{Strong} &
		148                       &
		144.80 & $\pm3.21$           &
		2.79 & $\pm0.05$            \\
		&
		                        &
		222                       &
		98.03 & $\pm3.32$        &
		2.51 & $\pm0.2$          \\
		&
		                      &
		296                     &
		78.14 & $\pm3.02$         &
		2.45 & $\pm0.1$          \\

		&
		&
		370                    &
		61.80 & $\pm3.35$         &
		2.81 & $\pm0.3$          \\
	    &
	    &
		444                     &
		52.72 & $\pm2.32$        &
		3 & $\pm0.8$         \\
	    &
	    &
		518                     &
		47.10 & $\pm3.54$        &
		3.5 & $\pm0.8$         \\
		\midrule
		%
		% \multirow{3}{*}{2}      &
		\multirow{12}{*}{Sort} &
		\multirow{6}{*}{Weak} &
		148                         &
		192.74 & $\pm3.21$         &
		3.87 & $\pm0.9$              \\
		&
		&
		222                      &
		204.44 & $\pm3.32$        &
		3.4 & $\pm1.2$              \\
		&
		&
		296                      &
		207.20 & $\pm4.02$        &
		3.85 & $\pm0.9$            \\

	         &
		&
		370                      &
		212.81 & $\pm3.35$         &
		2.59 & $\pm0.39$          \\

		&&
		444                     &
		215.05 & $\pm3.32$        &
		2.61 & $\pm0.88$          \\

		&&
		518                    &
		223.88 & $\pm4.54$          &
		3.23 & $\pm1.3$        \\
		\cmidrule{3-7}
		&
		\multirow{7}{*}{Strong} &
		148                         &
		125.53 & $\pm2.64$             &
		2.42 & $\pm0.8$              \\
		&
		&
		222                       &
		84.20 & $\pm2.64$            &
		2.37 & $\pm0.61$               \\
		&
		&
		296                      &
		63.76 & $\pm2.80$           &
		2.42 & $\pm0.5$           \\
		&&
		370                     &
		51.31 & $\pm3.18$           &
		2.65 & $\pm0.92$           \\
		&&
		444                     &
		44.46 & $\pm0.96$       &
		2.91 & $\pm0.8$        \\

		&&
		518                     &
		39.52 & $\pm3.98$       &
		3.5 & $\pm1.05$         \\
		\bottomrule
	\end{tabular}
\end{table}

\subsection{Join Operation Scalability}\label{subsec:exp-join}
\begin{figure}
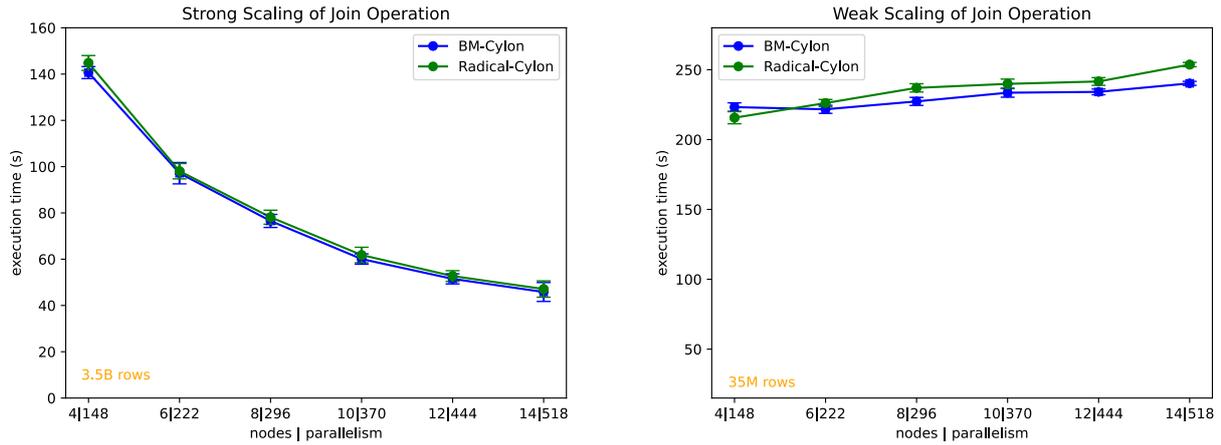

    \includesvg[width=0.5\linewidth]{Figure/rivanna-join-s-scaling}\hfill
    \includesvg[width=0.5\linewidth]{Figure/rivanna-join-w-scaling}
    \caption{Comparison of strong scaling(left) and weak scaling(right) performance of Bare-Metal and Radical-Cylon with join operation on \textbf{\textit{Rivanna}}. ~\texttt{execution time(s)} is calculated by running task for 10 iterations. The number of parallelism is calculated by nodes multiple by 37 cores per node}
    \label{fig:rivanna-join-sw-scaling}
\end{figure}

\begin{figure}
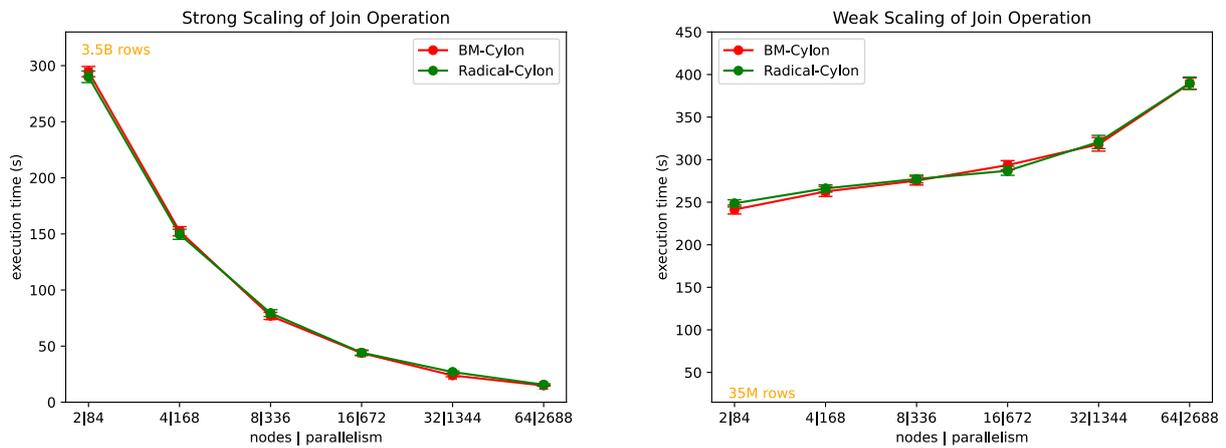

    \includesvg[width=0.5\linewidth]{Figure/summit-join-s-scaling}\hfill
    \includesvg[width=0.5\linewidth]{Figure/summit-join-w-scaling}
    \caption{Comparison of strong scaling(left) and weak scaling(right) performance of Bare-Metal and Radical-Cylon with join operation on \textbf{\textit{ORNL-Summit}}. ~\texttt{execution time(s)} is calculated by running the task for 10 iterations and it's used for a higher scalability test. The number of parallelism is calculated by nodes multiple by 42 cores per node.}
    \label{fig:summit-join-sw-scaling}
\end{figure}

The join weak scaling experiment is depicted in Fig.-\ref{fig:rivanna-join-sw-scaling} (right), \ref{fig:summit-join-sw-scaling} (right). 
%In an ideal scenario, the execution time for each job would remain constant, forming a flat line. However, due to system overhead, total execution time based on 10 iterations is considered for each task. 
Across all tests are performed 10 times with multiple parallelisms (a single rank is used for each parallel execution), and the total execution time ranges from 215 to 250 seconds for both bare-metal and RADICAL-Pilot Cylon executions on Rivanna. We got an overlapping error bar by increasing the number of workers in \textbf{\textit{ORNL-Summit}} with different configurations because of constant RP overheads (Fig.~\ref{fig:summit-join-sw-scaling} (right)).

As the number of ranks increases, a minor addition of execution time for allgather from all ranks becomes evident in the performance in the join weak scaling experiment. It's noteworthy that Radical-Cylon exhibits better performance with a lower number of ranks, particularly when it's below 200. Starting from 222 ranks and onwards, in join weak scaling, an average of 10 seconds of increasing is observed, which is deemed acceptable considering the benefit of achieving heterogeneity among multiple nodes in the HPC system. However, the error bar graph in both Cylon overlaps indicates we achieved similar execution times in a single pipeline. Similar trends can be observed in the join strong scaling operation. For strong scaling, where 3.5 billion rows are distributed among all ranks, the same 10 iterations are employed. The results, depicted in Fig.-\ref{fig:rivanna-join-sw-scaling} (left), \ref{fig:summit-join-sw-scaling} (left) demonstrate a significant reduction in execution time as the number of ranks increases for both BM-Cylon and Radical-Cylon implementations on Rivanna and ORNL-Summit.

With the increasing rank count, Radical-Cylon gradually closes the latency gap with BM-Cylon, showing only a marginal difference in latencies of total execution time. The error bar shows an identical performance with both experiments set up. This leveling of latencies can be attributed to the efficient scheduling and task distribution mechanisms employed by Radical-Cylon. This efficiency arises from the fixed allocation of rows among ranks for join operations and the utilization of a consistent table index for merging in distributed join operations. Consequently, the communication and aggregation overheads are constant in Radical-Cylon (in Table \ref{tab:a-b-exp_table}).

\subsection{Sort Operation Scalability}\label{subsec:exp-sort}
\begin{figure}
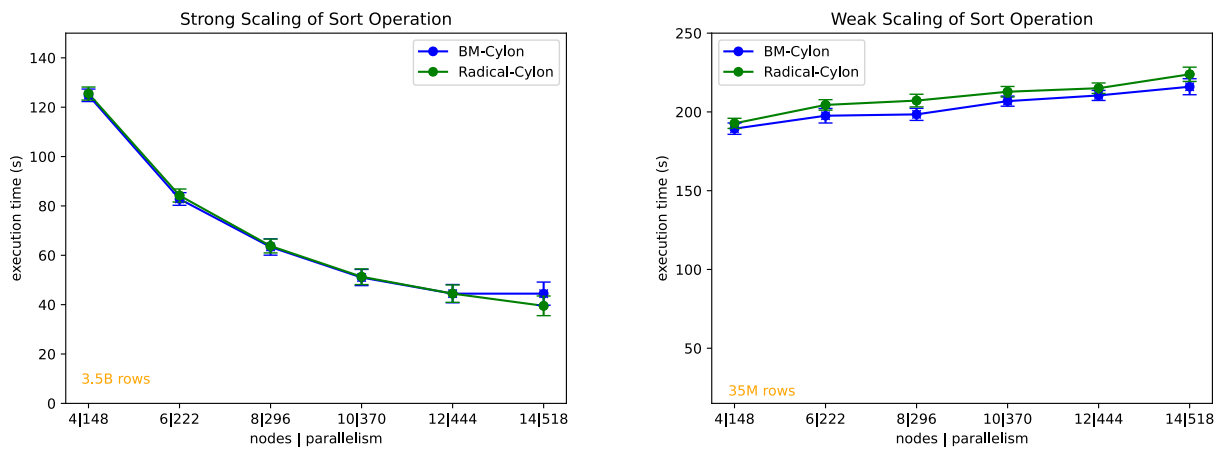

    \includesvg[width=0.5\linewidth]{Figure/rivanna-sort-s-scaling}\hfill
    \includesvg[width=0.5\linewidth]{Figure/rivanna-sort-w-scaling}
    \caption{Comparison of strong scaling(left) and weak scaling(right) performance of Bare-Metal and Radical-Cylon with sort operation on \textbf{\textit{Rivanna}}. ~\texttt{execution time(s)} is calculated by running task for 10 iterations. The number of parallelism is calculated by nodes multiple by 37 cores per node}
    \label{fig:rivanna-sort-sw-scaling}
\end{figure}

\begin{figure}
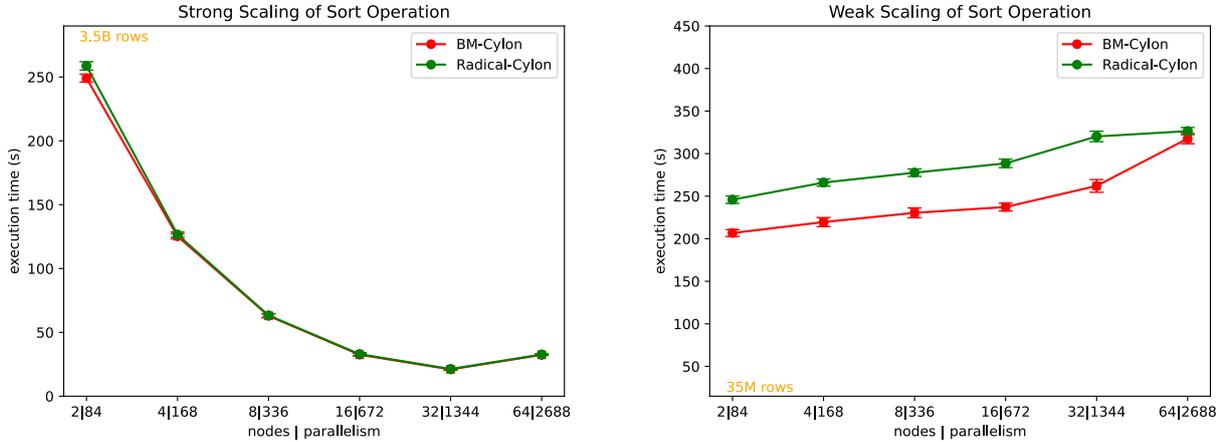

    \includesvg[width=0.5\linewidth]{Figure/summit-sort-s-scaling}\hfill
    \includesvg[width=0.5\linewidth]{Figure/summit-sort-w-scaling}
    \caption{Comparison of strong scaling(left) and weak scaling(right) performance of Bare-Metal and Radical-Cylon with sort operation on \textbf{\textit{ORNL-Summit}}. The number of parallelism is calculated by nodes multiple by 42 cores per node. Strong scaling with 2688 nodes takes a bit more time than with 1344 due to the lack of rows available for each worker and some workers go idle.}
    \label{fig:summit-sort-sw-scaling}
\end{figure}

The identical scaling configurations are applied to sorting operations, and a single rank is used in each parallel execution on Rivanna and ORNL-Summit (Fig.-\ref{fig:rivanna-sort-sw-scaling} (right), \ref{fig:summit-sort-sw-scaling} (right)). In Fig.-\ref{fig:rivanna-sort-sw-scaling} (right), for Rivanna, an average latency discrepancy of around 15 seconds is observed between the minimum rank count (148) and the maximum rank count (518) in the weak scaling experiment. This latency increase with higher rank numbers is anticipated, as it influences the data shuffling and merging stages within the distributed sorting process, thereby introducing additional overhead. Effective utilization of resources for communication and data partition is pivotal in influencing execution time. Remarkably, as the rank count increases, Radical-Cylon demonstrates enhanced performance and consistently narrows the gap with BM-Cylon. But with multiple iterations, we are getting an overlapping error bar that indicates similar performance with both metrics.

%The benefits of distributed sorting for large-scale datasets and its impact on results are vividly illustrated in the strong scaling sort operation. 
Partitioning a massive dataset across numerous nodes leads to a reduction in execution time. In the strong scaling sort operation, showcased in Fig.-\ref{fig:rivanna-sort-sw-scaling} (left) and \ref{fig:summit-sort-sw-scaling} (left), a tabular dataset containing 3.5 billion rows is partitioned among hundreds of ranks across various test runs. Each test run encompasses 10 iterations, and the execution time is utilized for graph plotting. The results unequivocally highlight that with 148 ranks, the total execution time amounts to 125 seconds, which diminishes to a mere 39.5 seconds with 518 ranks on the Rivanna cluster. Due to constant overheads in both scaling of the sort operation, the same behavior is observed in the ORNL-Summit.

Both RADICAL-Pilot and BM-Cylon approaches achieve closely comparable performance, differing by mere milliseconds in their total execution times, although, with multiple iterations, we are getting an overlapping error bar. However, distributed execution introduces a set of challenges encompassing the management of data distribution, navigation of communication overhead between nodes, and mitigation of potential node failures. These complexities are magnified with an increased number of nodes. That might happen in both BM and Radical Cylon.
%We introduced an increment of 74 ranks in each test set. Initially, the execution time exhibits a rapid decline, from 148 ranks to 222 ranks, resulting in a reduction from 125.5 to 84.2 seconds (approximately 33\%). However, this rate of reduction does not maintain its pace, tapering off to an 11\% reduction as the rank count increases from 444 to 518. 
Apart from the comparable performance, we see a constant overhead when using Radical-Cylon in strong and weak scaling operations (in Table \ref{tab:a-b-exp_table}) despite of increasing parallelism.

\begin{figure}[htpb]
    \begin{center}
    \includesvg[width=0.7\linewidth]{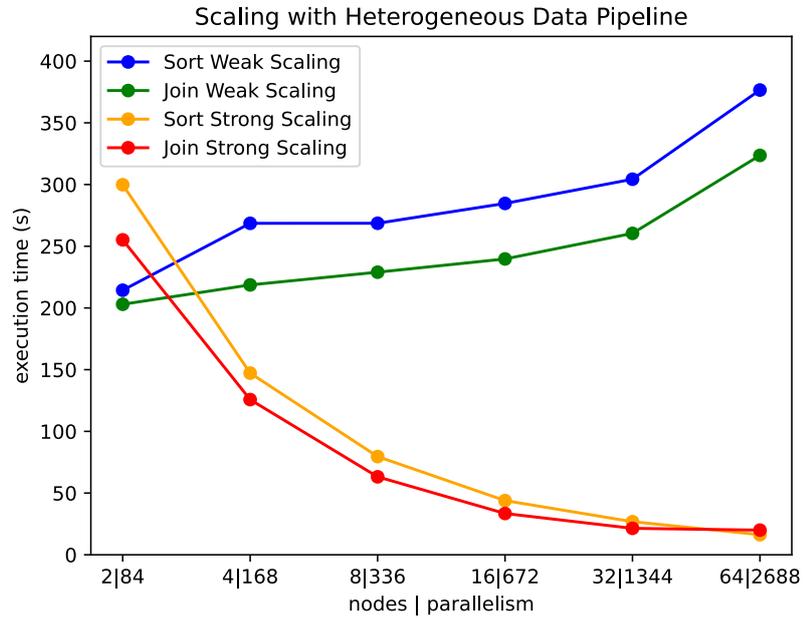}
    \end{center}
    \caption{Heterogeneous Executions with sort and join strong and weak scaling(4) operations on \textbf{\textit{ORNL-Summit}}. Strong scaling with 2688 nodes takes a bit more time than with 1344 due to the lack of rows available for each worker and some workers go idle.}
    \label{fig:summit-heterogeneous-scaling}
\end{figure}

%\subsection{Overheads of RP-Cylon}

\subsection{Benchmarking with multiple data pipeline}\label{subsec:exp-homogeneous-cylon}

The heterogeneous data pipeline is used on the ORNL-Summit clusters, involving multiple scaling benchmarks. 
%In the initial experiment, the goal is to validate the task-based execution environment through the repetition of a single distributed join operation three times. This aims to observe how resources are allocated and redistributed to subsequent tasks. Since the same task is executed, managing data in memory across multiple workers in different CPUs does not present complexity. To assess the viability of outcomes concurrently, various operations with multiple data pipelines are implemented. 
Four distinct scaling operations, namely Sort and Join weak scaling (WS), are configured with 35 million rows in each worker, while strong scaling is executed with 3.5 billion rows. Six different experiments are conducted with CPU counts ranging from 84 to 2688. Each experiment is iterated 10 times in a single run. We gauge the total execution time (in seconds) against the number of nodes or parallelism and illustrate the results in Fig.-\ref{fig:summit-heterogeneous-scaling}. 
%All four different operations are executed, yielding the anticipated results.

In the case of weak scaling for the sort and join operations, there is a gradual increase in execution time to compile results for generating a global table. as the number of CPUs rises. Similarly, with strong scaling operations, where 3.5 billion rows are distributed among multiple workers, performance improves as the number of workers increases, resulting in a significant reduction in execution time. This experiment validates the achievement of a scalable model using the proposed task-based execution framework.

%\blindtext
\begin{figure}
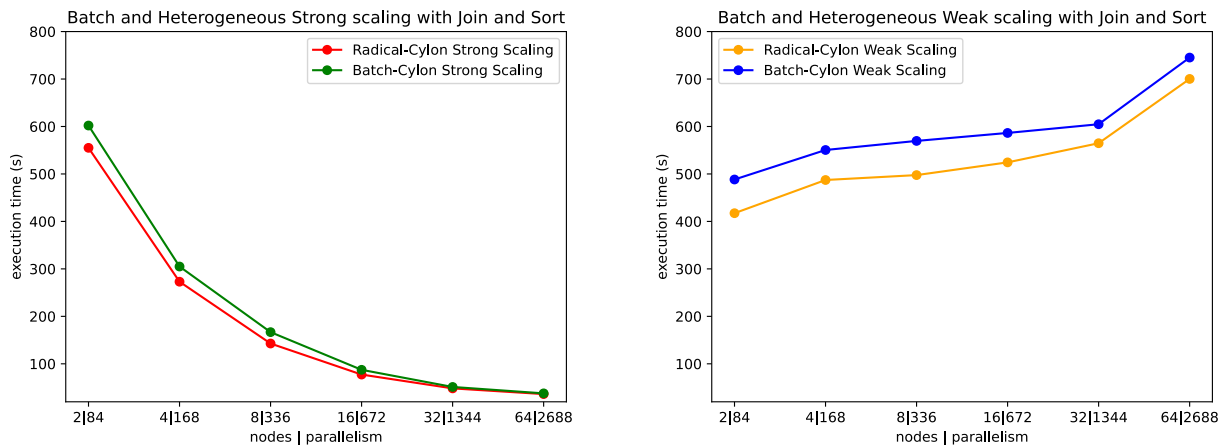

\includesvg[width=.5\linewidth]{Figure/summit-heterogeneous-s-comp}\hfill
\includesvg[width=.5\linewidth]{Figure/summit-heterogeneous-w-comp}
\caption{Comparison of strong scaling(left) and weak scaling(right) performance of Heterogeneous and Batch executions on \textbf{\textit{ORNL-Summit}}. ~\texttt{execution time(s)} is calculated by running task for 10 iterations. Batch execution time for join and sort is calculated separately from two batch outputs}
\label{fig:summit-heterogeneous-sw-comp}
\end{figure}
%\blindtext

However, the core premise of our argument faces a potential challenge if the proposed design cannot surpass the performance of Batch execution while ensuring minimal resource utilization. In the Batch execution model, join and sort operations are configured through an LSF-based script on the ORNL-Summit cluster, running in parallel. 
%Each script independently requests a separate allocation of hardware resources. The system allocates available resources based on walltime separately for the two distinct operations. For instance, if we request 2 nodes with 42 cores per node, during batch execution, the join and sort operations will each receive a separate allocation of 84 CPUs to execute in parallel. The downside lies in the additional overhead introduced by system wait time during resource scheduling and allocation. 
Each operation lacks control over the hardware resources of the other operation, even if some workers finish their tasks, introducing a potential inefficiency in resource usage.

\begin{figure}[htpb]
    \begin{center}
    \includesvg[width=0.7\linewidth]{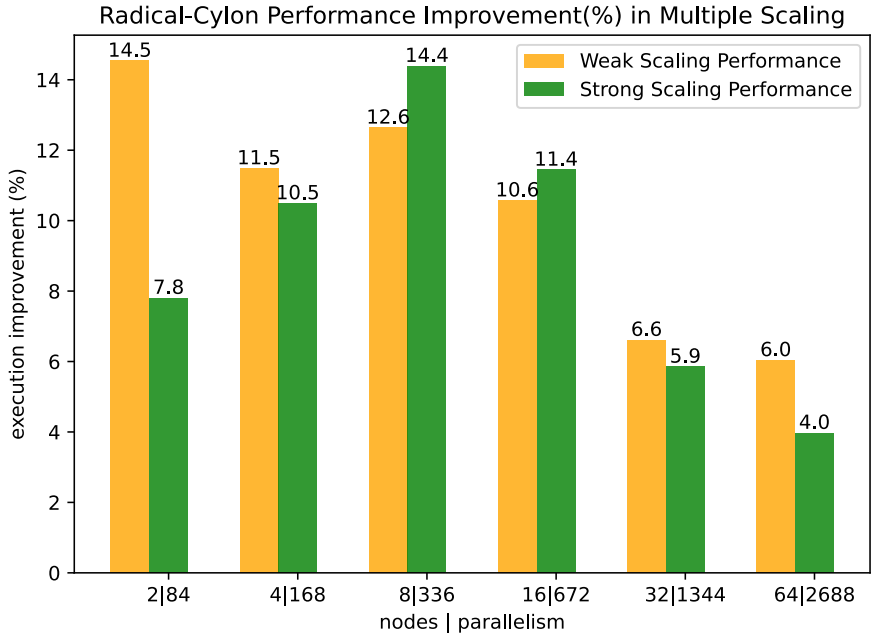}
    \end{center}
    \caption{Radical-Cylon performance improvement with scaling operation on \textbf{\textit{ORNL-Summit}}. The number of nodes for batch execution is generated by the requested resource as the system allocates resources separately}
    \label{fig:summit-heterogeneous-performance}
\end{figure} 

In the context of the heterogeneous scaling operation, the join and sort processes are treated as distinct tasks within a single execution. Consequently, when any worker completes their task, the released resources become available to others. For weak scaling join and sort operations (depicted in Fig.-\ref{fig:summit-heterogeneous-sw-comp} (right)), 84 CPUs are efficiently allocated, and resource release is effectively managed, enabling both tasks to conclude in 417.33 seconds. In contrast, under the batch execution model, the same amount of CPUs are allocated separately for the join and sort processes, consuming a total of 488.33 seconds to execute both tasks, despite resource allocation considerations.

The same efficiency is observed in strong scaling join and sort operations for both heterogeneous and batch execution, as illustrated in Fig.-\ref{fig:summit-heterogeneous-sw-comp} (left). Radical-Cylon achieves comparable or improved execution times while utilizing the same resources for the two tasks, thanks to additional optimizations in separate resource utilization and constant RP overheads. To provide a comprehensive overview of the performance evaluation between heterogeneous and batch execution, we have plotted a bar graph (Fig-\ref{fig:summit-heterogeneous-performance}) depicting performance improvement with multiple configurations. Radical-Cylon consistently outperforms batch execution by 4-15\% in various configurations of scaling operations. This underscores the effectiveness of Radical-Cylon in achieving superior performance with optimized resource utilization. 

\subsection{Discussions}
\label{sec:discussions}

The results in \ref{subsec:exp-join},~\ref{subsec:exp-sort} show that Radical-Cylon scales efficiently with very small overheads and achieves similar performance in single-task execution.  It also outperforms batch processing of BM-Cylon in ~\ref{subsec:exp-homogeneous-cylon} with heterogeneous execution. Radical-Cylon overheads presented in Table~\ref{tab:a-b-exp_table} show that Radical-Cylon takes an average of 3.4 seconds to construct an~\texttt{MPI-Communincator} with 518 ranks which are marginal compared to the total execution time and the size of the experiments. It shows impeccable scalability, particularly when the number of CPUs is less than or equal to 2688 on ORNL-Summit. The Raptor module, responsible for resource allocation and scheduling, unfortunately, encountered challenges in allocating resources for Cylon. We are working consistently with  OLCF support team to fix the issue.

Cylon performance is measured with a data frame execution runtime. A collection of data frame operators can be arranged in a directed acyclic graph (DAG). Execution of this DAG can further be improved by identifying independent branches of execution and executing such independent tasks parallelly. Additionally, each of these tasks themselves is Bulk Synchronize Parallel (BSP). Radical-Cylon allows us to control the parallelism of these BSP tasks. 
In the future, Cylon is planning to add an optimizer based on the data frame DAG. One aspect is traditional query optimization, similar to SQL query optimization, which is orthogonal to the scheduling mechanism. Another important metric would be scheduling overhead distribution of the underlying scheduling environment. This forward-looking initiative aims to enhance the efficiency and performance of data processing within the context of machine/deep learning tasks.

Radical-Cylon has been meticulously crafted to address a unified execution setting encompassing both CPUs and GPUs, catering to the needs of multiple data pipelines. While integrating CPUs and GPUs within a single task does introduce computational intricacies owing to the foundational structure of Cylon, the concept of heterogeneous execution remains viable. This involves employing distinct groups of ranks equipped with specialized memory allocated either on CPUs or GPUs, enabling the concurrent utilization of these processing units. However, we limit our experiments to CPU clusters only due to dependencies on CUDA-aware MPI along with supported frameworks of ORNL-Summit and Rivanna cluster, which is in the process of being addressed.

The design of RADICAL-Pilot aims to support an extensive spectrum of meticulous resource management policies. Our initial focus has revolved around establishing multiple data pipelines and executing distributed operations in the form of functions. Here Cylon plays an important part by providing distributed execution. Looking ahead to more complex multi-tenancy scenarios, RADICAL-Pilot must proficiently manage a diverse range of resource types, including not only device and host memory but also network bandwidth. The Master-Worker raptor model employed by RADICAL-Pilot provides the system with a robust capability to monitor available resources and allocate them on a large scale. Our future plans involve exploring common multi-tenancy requirements such as prioritization, performance segregation, and resource tracking. Importantly, the timeframe for these endeavors is considerably shorter than prior efforts yet will encompass significantly larger pools of resources that will help to implement a seamless ML/DL pipeline.

\section{Conclusions}
\label{sec:conclusion}

RP achieves parity with the cutting-edge multi-execution design in today's data engineering execution landscape, which predominantly employs an SPMD approach. This compatibility extends to multi-execution setups using SLURM-SRUN on top of Cylon, as evidenced in our evaluation section. RP effectively tackles the intricacies of resource management and the execution of diverse data pipelines. Notably, RADICAL-Pilot attains performance levels comparable to Bare-Metal Cylon across various distributed operations.

Concurrently, RP revolutionizes the execution model of Cylon programs, consolidating user code under a single execution framework. This transformation introduces a centralized resource management and scheduling framework that interfaces between the client and cluster nodes. The outcome of this unified execution model is enhanced user access to more comprehensive computation patterns. Our micro-benchmarks substantiate the effective interleaving of client workloads and streamlined pipelined execution, firmly establishing the efficiency and adaptability of the system with minimal overhead. Furthermore, the resource management and scheduling layer facilitates the reimplementation of cluster management policies, such as multi-execution sharing and virtualization, tailored specifically to the demands of ML and BigData workloads.

%The agent-based model underpinning RP harmonizes diverse dependencies, encompassing message passing while inherently supporting pipelining. This serves as a distinctive mechanism for the runtime of distributed execution. 

%Presently, our focus lies on the ongoing deployment of the system design, engineering, and integration onto the Summit supercomputer. This strategic move is aimed at further benchmarking and validating the scalability and performance robustness of RP under demanding conditions.

\section*{Acknowledgments}
We gratefully acknowledge the support from the Department of Energy and National Science Foundation through DE-SC0023452, NSF 1931512, and NSF 2103986 grants.

\bibliographystyle{splncs04}
\bibliography{conference_101719}
\end{document}